\RequirePackage{ifpdf}
\ifpdf 
\documentclass[pdftex]{sigma}
\else
\documentclass{sigma}
\fi

\numberwithin{equation}{section}

      \def\cH{{\cal H}}

\def\cS{{\cal S}}

\def\fS{{\mathfrak S}}

\def\fm{{\mathfrak m}}

\def\fs{{\mathfrak s}}

\newcommand{\CC}{{\mathbb C}}

\newcommand{\half}{\frac{1}{2}}

\begin{document}

\allowdisplaybreaks

\renewcommand{\thefootnote}{$\star$}

\renewcommand{\PaperNumber}{006}

\FirstPageHeading

\ShortArticleName{Coordinate Bethe Ansatz for Spin $s$ XXX Model}

\ArticleName{Coordinate Bethe Ansatz for Spin $\boldsymbol{s}$ XXX Model\footnote{This paper is a
contribution to the Proceedings of the International Workshop ``Recent Advances in Quantum Integrable Systems''. The
full collection is available at
\href{http://www.emis.de/journals/SIGMA/RAQIS2010.html}{http://www.emis.de/journals/SIGMA/RAQIS2010.html}}}

\Author{Nicolas~CRAMP\'E~$^{\dag\ddag}$, Eric RAGOUCY~$^\S$ and Ludovic ALONZI~$^\S$}

\AuthorNameForHeading{N.~Cramp\'e, E.~Ragoucy and L.~Alonzi}

\Address{$^\dag$~Universit\'e Montpellier 2, Laboratoire Charles Coulomb UMR 5221,\\
\hphantom{$^\dag$}~F-34095 Montpellier, France}

\Address{$^\ddag$~CNRS, Laboratoire Charles Coulomb UMR 5221, F-34095 Montpellier, France}
\EmailD{\href{mailto:ncrampe@um2.fr}{ncrampe@um2.fr}}

\Address{$^\S$~LAPTh, CNRS and Universit{\'e} de Savoie, 9 chemin de Bellevue,\\
\hphantom{$^\S$}~BP 110, 74941, Annecy-Le-Vieux Cedex, France}
\EmailD{\href{mailto:eric.ragoucy@lapp.in2p3.fr}{eric.ragoucy@lapp.in2p3.fr}}

\ArticleDates{Received September 06, 2010, in f\/inal form January 05, 2011;  Published online January 12, 2011}

\Abstract{We compute the eigenfunctions and eigenvalues of the periodic
integrable spin~$s$
XXX model using the coordinate Bethe ansatz. To do so, we compute
explicitly the Hamiltonian of the model. These results generalize
what has been obtained for spin $\half$ and spin 1 chains.}


\Classification{81R12; 17B80}

\renewcommand{\thefootnote}{\arabic{footnote}}
\setcounter{footnote}{0}

\section{Introduction}

The resolution of  Heisenberg spin chain \cite{heis}
was initiated in H. Bethe's seminal paper \cite{bethe}
where he used a method called now coordinate Bethe ansatz.
Since this work, several new methods appeared: algebraic Bethe
ansatz \cite{kusk,tafa},
functional Bethe ansatz (or separation of variables) \cite{skl} or
analytical Bethe ansatz \cite{reshe}.
These more elaborated techniques allowed one to go further: new
solvable models
have been discovered and new results have been computed such as
correlation functions. As a consequence, the coordinate Bethe ansatz
was neglected.
However, this method is the simplest one and gives a very ef\/f\/icient
way to construct
explicitly eigenfunctions, but it is believed that it works only for
simple models.
In this note, we show that actually it can be applied also to
more complicated models
as the spin $s$ XXX model.

This paper is organized as follows. In Section~\ref{sec:ham},
we compute the
Hamiltonian
of the spin $s$ chain we want to solve. To our knowledge, the
explicit form of the entries of the Hamiltonian
are written for the f\/irst time. We also compte the $su(2)$ symmetry
algebra and the pseudo-vacuum, a particular (reference) eigenstate.
In Section~\ref{sec:spins}, we present the
coordinate Bethe ansatz
and get the Bethe equations obtained previously by the algebraic or
analytical Bethe ansatz.
We conclude, in Section~\ref{conclu}, on the advantages of this method
and on open problems.

\section[Integrable periodic spin $s$ chain]{Integrable periodic spin $\boldsymbol{s}$ chain}\label{sec:ham}

\subsection[Hamiltonian of the spin $s$ chain]{Hamiltonian of the spin $\boldsymbol{s}$ chain}

The Hamiltonian of the periodic integrable spin~$s$ chain has been
computed in~\cite{krs}
thanks to a~fusion procedure. This Hamiltonian has been expressed as
a polynomial of the invariant
of~$su(2)$ (see for example also \cite{fad} for a review).
For our purpose, we need to give an explicit expression of the
 Hamiltonian entries.
Namely, the Hamiltonian is the following matrix acting on
$(\CC^{2s+1})^{\otimes L}$
\begin{equation}
 \cH=\sum_{j=1}^L h_{j,j+1}
\label{eq:Htot}
\end{equation}
with the periodic condition $L+1=1$ and the subscript $(j,j+1)$ stands
for the two spaces where the $(2s+1)^2\times (2s+1)^2$-matrix $h$
acts non-trivially. We choose to enumerate
the basis of $\CC^{2s+1}$ as follows:  $|s\rangle$, $|s-1\rangle$, \dots,
$|-s\rangle$, where $|m\rangle\equiv|s,m\rangle$ denotes\footnote{Let
us stress that since all the sites have spin $s$, we do not mention it,
and write only the $\fs^z$ value of the states.} the spin $s$
state with $\fs^z$-component equals to $m$.

The non-vanishing entries of the matrix $h$ may be parameterized by
three integer parameters $m_1$, $m_2$, $n$
\begin{gather*}
h = \sum_{m_{1},m_{2}=0}^{2s}\
\sum_{n= -\min(m_1,2s-m_2)}^{\min(m_2,2s-m_1)}
\ \beta_{m_1,m_2}^n\
| s-m_1-n\rangle\langle s-m_1|\otimes
| s-m_2+n\rangle\langle s-m_2|  .
\end{gather*}
From the results found in \cite{krs}, one may prove that
$\beta_{m_1,m_2}^n$
can be factorised as:
\begin{gather}
\beta_{m_{1},m_{2}}^n  =
\frac{(-1)^{n-1}}{n}
\sqrt{\frac{\left(\begin{array}{c} M_{1}+n \\ M_{1}\end{array}\right)
\
\left(\begin{array}{c} M_{2} \\ n\end{array}\right)}
{\left(\begin{array}{c} 2s-M_{1} \\ n\end{array}\right)\
\left(\begin{array}{c} 2s-M_2+n \\ n\end{array}\right)}}\qquad
\mbox{for} \ \ n>0   ,
\label{eq:beta-facto}
\end{gather}
where we have introduced the notation
\begin{gather}
M_{1}=\min(m_{1},2s-m_{2}) \qquad \mbox{and} \qquad M_{2}=\min(m_{2},2s-m_{1}).
\label{eq:M1M2}
\end{gather}
The remaining $\beta$'s are given by the relations:
\begin{gather*}
\beta_{m_1,m_2}^0
 =
-\sum_{\ell=0}^{M_1-1}\frac{1}{2s-\ell}
-\sum_{\ell=0}^{M_2-1}\frac{1}{2s-\ell}  ,
\\
\beta_{m_{2},m_{1}}^{-n}  =  \beta_{m1,m_{2}}^n
 .
\end{gather*}

\subsection[$su(2)$ symmetry]{$\boldsymbol{su(2)}$ symmetry}

At each site, we have a spin $s$ representation, and the expression
of the $su(2)$ generators in this representation reads:
\begin{gather*}
\fs^-  =
\sum_{n=-s+1}^s\sqrt{(s+n) (s-n+1)}  |n-1\rangle\langle n| ,
\\
\fs^+  =
\sum_{n=-s}^{s-1}
\sqrt{(s-n)(s+n+1)}  |n+1\rangle\langle n| ,
\\
\fs^z  =  \sum_{n=-s}^{s}n  |n\rangle\langle n| .
\end{gather*}
They obey
\begin{gather*}
[\fs^+,\fs^-]=2\fs^z ,\qquad  [\fs^z,\fs^\pm]=\pm \fs^\pm,\qquad
c_{2}=(\fs^z)^2+\half(\fs^+\fs^-+\fs^-\fs^+)=s(s+1).
\end{gather*}
We will note $\fs^{\alpha}_{j}$, $\alpha=z,\pm$ and $j=1,\ldots,L$, the
generators acting on site $j$. Let us stress that these local
operators do not commute with the Hamiltonian  $\cH$ given in~(\ref{eq:Htot}).
However, there is a global $su(2)$ symmetry. The generators
of this $su(2)$ symmetry take the form
\begin{gather*}
\cS^z=\sum_{j=1}^{L} \fs^z_{j}\qquad \mbox{and}\qquad
\cS^\pm=\sum_{j=1}^{L} \fs^\pm_{j}.
\end{gather*}
They obey the $su(2)$ commutation relations $[\cS^z,\cS^\pm]= \pm  \cS^\pm$,
$[\cS^+,\cS^-]=2 \cS^z$. Remark that the Casimir operator
$C_{2}=(\cS^z)^2+\half(\cS^+ \cS^-+\cS^- \cS^+)$, although central, is not
proportional to the identity, since we are considering the tensor product
of $L$ spin $s$ representations, a reducible representation.

It is a simple calculation to show that
$[\fs^{\alpha}_{j}+\fs^{\alpha}_{j+1},h_{j,j+1}]=0$, $\alpha=z,\pm$.
It amounts to check the following recursion relations on the coef\/f\/icients
$\beta_{m_{1},m_{2}}^n$:
\begin{gather*}
\sqrt{(m_1+1)(2s-m_1)} \beta_{m_1+1,m_2}^{n}=
\sqrt{(2s+n-m_2+1)(m_2-n)} \beta_{m_1,m_2}^{n+1}
\nonumber\\
\qquad{} -\sqrt{(m_2+1)(2s-m_2)} \beta_{m_1,m_2+1}^{n+1}
+\sqrt{(2s-n-m_1)(n+m_1+1)} \beta_{m_1,m_2}^{n},
\\
\sqrt{m_1(2s-m_1+1)} \beta_{m_1-1,m_2}^{n}=
\sqrt{(2s+n-m_2)(m_2-n+1)} \beta_{m_1,m_2}^{n-1}
\nonumber\\
\qquad{} -\sqrt{m_2(2s-m_2+1)} \beta_{m_1,m_2-1}^{n-1}
+\sqrt{(2s-n-m_1+1)(n+m_1)} \beta_{m_1,m_2}^{n}.
\end{gather*}
Hence, $\cS^\alpha$ commutes with $\cH$.

Due to this $su(2)$ symmetry, the wave functions can be characterized
by their energy, their spin and their $\cS^z$ component.
In other words,
one can diagonalize the Hamiltonian in a sector where the operators
$\cS^z$
has a f\/ixed value $S^z=Ls-m$. This is done in the next section.

\subsection{Pseudo-vacuum and pseudo-excitations}

We wish to present the construction of the Hamiltonian eigenfunction
in the framework of coordinate Bethe ansatz for spin $s$. The
spin $s=\half$ case
 is the Heisenberg chain, solved in \cite{bethe}. It
gave the name
to the method. For the case $s=1$, the method has been generalized
in \cite{LimaS}.

The f\/irst step of the coordinate Bethe ansatz \cite{bethe} consists
in f\/inding a particular
eigenvector, called the pseudo-vacuum, for the Hamiltonian. It is
usually chosen as the vector with the highest spin. In the present
case, it
is the unique vector in the sector $S^z=Ls$:
\begin{gather*}
 |\varnothing\rangle= |s\rangle \otimes |s\rangle \dots \otimes|s\rangle
.
\end{gather*}
Using the explicit forms for the $\beta$'s, we get $h_{12}\,|s\rangle
\otimes |s\rangle=0$. Thus,
$|\varnothing\rangle$ is a $\cH$-eigenvector with vanishing
eigenvalue.

The second step consists in adding pseudo-excitations. These pseudo-excitations
are not physical excitations (hence the name pseudo-excitation). They are
obtained by acting with a creation
 operator $e^-_{j}$, conjugated to $\fs^-_j$,
on the pseudo-vacuum $|\varnothing\rangle$.
Let us remark that this operator in this f\/inite representation
does not satisfy $(e^-)^2=0$ but rather $(e^-)^{2s+1}=0$.
This  explains the supplementary dif\/f\/iculties to deal with
$s>\half$.
Indeed, in the case $s=\half$,
no more than one pseudo-excitation can be at the same site: we have strict
exclusion.
In the general case of spin $s$, we have a weaker exclusion. More
precisely,
 we can have up to $2s$ pseudo-excitations at the same site. This
behavior appears already for $s=1$.

\section[Coordinate Bethe ansatz for general spin $s$]{Coordinate Bethe ansatz for general spin $\boldsymbol{s}$}\label{sec:spins}

We def\/ine a state in the sector $S^z=Ls-m$, for $1\leq x_1\leq x_2\leq \dots
\leq x_m\leq L$
\begin{gather}
\label{eq:vm}
 |x_1,x_2,\dots,x_m\rangle=
e^-_{x_1}e^-_{x_2}\cdots e^-_{x_m}|\varnothing \rangle ,
\end{gather}
where $e^-$ is conjugated to $\fs^-$:
\begin{gather*}
e^-  =
\sum_{n=-s+1}^s\sqrt{\frac{s+n}{s-n+1}}\ |n-1\rangle\langle n|
 =  g  \fs^-  g^{-1} \qquad \mbox{with} \quad g=\sum_{n=-s}^s
\frac{(2s)!}{(s-n)!}  |n\rangle\langle n| .
\end{gather*}
This choice for $e^-$ is for later convenience.
The set of such non-vanishing vectors
(i.e.\ $x_{j+2s}>x_j$ for $1\leq j \leq L-2s$)
provides a basis
for this sector. As already noticed, contrarily to
 the usual case
($s=\half$), several (up to $2s$) pseudo-excitations at the same site are
allowed, that is to say, some $x_{j}$'s can be equal.
The restriction that no more than $2s$ particles are
on the same site is implemented by the fact that
$(e^-)^{2s}=0$.
Using the explicit form of $e^-$, we can rewrite the
excited states as follows:
\begin{gather*}
 |x_1,x_2,\dots,x_m\rangle =
\alpha_{m_1}\cdots\alpha_{m_k}
\underbrace{|s\rangle\otimes\cdots\otimes|s\rangle}_{x_1-1}
\otimes|s-m_1\rangle\otimes
\underbrace{|s\rangle\otimes\cdots\otimes|s\rangle}_{x_{m_{1}+1}-x_{m_1}-1}
\otimes|s-m_2\rangle\otimes
\cdots,
\end{gather*}
where $m_{j}$ is the number of times $x_{j}$ appears and
\begin{gather*}
 \alpha_m=\sqrt{
\left( \begin{array}{c}
2s\\
m
\end{array}
\right)} ,
\end{gather*}
where $\left(\begin{array}{c}
z\\ k \end{array} \right)=
\frac{z(z-1)\cdots(z-k+1)}{k!}$ is the binomial coef\/f\/icient. Let us
remark that, if $m_j>2s$ the vector $|s-m_j\rangle$ has no meaning
but the normalization $\alpha_{m_j}$ vanishes.

Any eigenvector in the sector $S^z=Ls-m$ is a linear combination of
the vectors (\ref{eq:vm}).
Then, let us introduce the vector
\begin{gather*}
 \Psi_{m}=\sum_{x_1\leq x_2\leq \cdots \leq x_m}a(x_1,x_2,\dots,x_m)
|x_1,x_2,\dots,x_m\rangle ,
\end{gather*}
where $a(x_1,x_2,\dots,x_m)$ are complex-valued functions to be
determined.
As in the case of $s=\half$, we assume a plane wave decomposition
for these functions (Bethe ansatz)
\begin{gather*}
a(x_1,\dots,x_m)=\sum_{P\in \fS_m} A_P(\textbf{k})  \exp
\big\{i(k_{P1}x_1+\cdots+k_{Pm}x_m)\big\} ,
\end{gather*}
where $\fS_m$ is the permutation group of $m$ elements and
$A_P(\textbf{k})$ are functions on the symmetric group algebra
depending on some parameters $k$ which will
be specif\/ied below\footnote{In the following, to lighten the
presentation, the
\textbf{k}-dependence will not be written explicitly.}.
Using the fact that the states $|x_1,x_2,\dots,x_m\rangle$ form a basis,
we can project the eigenvalue equation
\begin{equation}\label{eq:vp}
\cH \Psi_{m}=E \Psi_{m}
\end{equation}
on these dif\/ferent basis vectors to determine the $A_P(\textbf{k})$ parameters.

 Since $\cH$ is a sum of operators acting on two neighbouring sites
 only, one has to single out the cases where the $x$'s obey the
 following constraints:
\begin{itemize}\itemsep=0pt
\item  all the $x_{j}$'s are far away one from each other and are not on the
boundary sites 1 and $L$ (this case will be called generic),
\item $x_{j}+1=x_{j+1}$ for some $j$,
\item  $x_{j}=x_{j+1}$ for some $j$,
\item  $x_{j}=x_{j+1}=\cdots=x_{j+m_{1}}$ and
$x_{j}+1=x_{j+m_{1}+1}=\cdots=x_{j+m_{1}+1+m_{2}}$ for some positive
integers $m_{1}$ and $m_{2}$,
\item $x_{1}=1$, or $x_{m}=L$.
\end{itemize}
As the eigenvalue problem is a linear problem, it is
enough to treat the cases where at most one of the particular cases
appears: more complicated cases just appear as superposition of these
`simple' cases.

{\bf Projection on $\boldsymbol{|x_1,x_2,\dots,x_m\rangle}$ with
$\boldsymbol{x_{j}+1<x_{j+1}}$, $\boldsymbol{\forall\, j}$, $\boldsymbol{x_1>1}$ and $\boldsymbol{x_m<L}$.}
As usual, we start by projecting (\ref{eq:vp}) on a generic vector
$|x_1,x_2,\dots,x_m\rangle$.
This leads to
\begin{gather*}
\sum_{P\in \fS_m}\!\!A_P\Bigg(\sum_{j=1}^m
(\beta_{1,0}^0+\beta_{0,1}^0+\beta_{1,0}^{-1}
e^{ik_{Pj}}+\beta_{0,1}^1 e^{-ik_{Pj}})-E\Bigg)\!
\exp (i(k_{P1}x_1+\cdots+k_{Pm}x_m))=0
\end{gather*}
which must be true for any choice of generic $x$'s.
Therefore, we get for the energy (using the explicit forms of the
$\beta$'s given in Section~\ref{sec:ham})
\begin{gather*}
 E=-\frac{1}{2s}\sum_{j=1}^m\big(2-e^{ik_j}-e^{-ik_j}\big) .
\end{gather*}
After the change of variable
\begin{gather}
e^{ik_j}=\frac{\lambda_j+is}{\lambda_j-is} ,
\label{eq:chgV}
\end{gather}
the energy becomes
\begin{gather*}
 E=-\sum_{j=1}^m\frac{2s}{\lambda_j+s^2} .
\end{gather*}
This form for the energy is the one obtained by algebraic Bethe
ansatz \cite{ABAs}.

{\bf Projection on $\boldsymbol{|x_1,x_2,\dots,x_m\rangle}$ with
$\boldsymbol{x_{j}+1=x_{j+1}}$ for some $\boldsymbol{j}$.}
Let us consider now the projection of (\ref{eq:vp})
 when two pseudo-excitations
are nearest neighbours. Using the form of the energy previously
found, we get
\begin{gather*}
\sum_{P\in \fS_m}A_P\big(\beta_{1,1}^0-\beta_{1,0}^0-\beta_{0,1}^0
+(\beta_{1,1}^{-1}\alpha_2-\beta_{1,0}^{-1})e^{ik_{Pj}}
+(\beta_{1,1}^1\alpha_2-\beta_{0,1}^1)
e^{-ik_{P(j+1)}}\big)\\
\qquad{}\times e^{i(\dots k_{Pj}x_j+k_{P(j+1)}(1+x_j)\dots)}
=0.
\end{gather*}
This equation is trivially satisf\/ied for $s>\half$ since using
explicit values we f\/ind
$\beta_{1,1}^0-\beta_{1,0}^0-\beta_{0,1}^0=0$,
$\beta_{1,1}^{-1}\alpha_2-\beta_{1,0}^{-1}=0$ and
$\beta_{1,1}^1\alpha_2-\beta_{0,1}^1=0$.
For the case $s=\half$, we f\/ind a constraint between $A_P$ and
$A_{PT_j}$ (where $T_j$ is the transposition of $j$ and $j+1$).
Explicitly, it is given by (\ref{eq:cS}) with $s=\half$.

{\bf Projection on $\boldsymbol{|x_1,x_2,\dots,x_m\rangle}$ with
$\boldsymbol{x_{j}=x_{j+1}}$ for some $\boldsymbol{j}$.}
For $s>\half$, we must also consider the case when several particles
are on the same site. Def\/ining~$S_j$, the shift operator adding $1$ to the $j^{\rm th}$ variable,
we get the following relation, when two particles are on the same
site
\begin{gather}
\frac{1}{2s(1-2s)}
(1+S_i^{-1}S_{i+1}^{-1})
\big(S_iS_{i+1}+(2s-1)S_i-(2s+1)S_{i+1}+1\big)
 a(\dots,x_i,x_i,\dots)=0\nonumber
  \\
\qquad{}\Rightarrow \quad
\big(S_iS_{i+1}+(2s-1)S_i-(2s+1)S_{i+1}+1\big)
 a(\dots,x_i,x_i,\dots)=0 .\label{eq2s}
\end{gather}

Using the plane waves decomposition,
we get the following constraint
\begin{gather}
\label{eq:cS}
 A_{PT_j}=\sigma\big(e^{ik_{Pj}},e^{ik_{P(j+1)}}\big)A_P ,
\end{gather}
where $T_j$ is the transposition of $j$ and $j+1$, and we have
introduced the scattering matrix
\begin{gather}
 \sigma(u,v)=-\frac{uv+(2s-1)u-(2s+1)v+1}
{uv+(2s-1)v-(2s+1)u+1} .
\label{eq:Smat}
\end{gather}
As in the case $s=\half$, relation (\ref{eq:cS}) allows us to express all the
$A_P$'s in terms of only one, for instance $A_{\rm Id}$
(where $\rm Id$ is the identity of $\fS_m$). More precisely, one expresses $P\in \fS_m$
as a~product of $T_i$, and then uses (\ref{eq:cS}) recursively to express $A_P$ in terms of
$A_{\rm Id}$. At this point, one must take into account
that the expression of $P$ in terms of $T_i$ is not unique,
because of the relations
\begin{gather*}
 T_i^2={\rm Id},\qquad [T_j,T_i]=0 \quad (|j-i|>1)\qquad \mbox{and} \qquad T_iT_{i+1}T_i=T_{i+1}T_{i}T_{i+1} .
\end{gather*}
Therefore, for the construction to be consistent, the function $\sigma$
has to satisfy the relations
\begin{gather*}
  \sigma(u,v) \sigma(v,u)=1 ,\qquad [\sigma(u,v) , \sigma(w,z)]=0 ,\\
 \sigma(u_1,u_2) \sigma(u_1,u_3) \sigma(u_2,u_3)=
\sigma(u_2,u_3) \sigma(u_1,u_3) \sigma(u_1,u_2) .
\end{gather*}
By direct computation, we can show that (\ref{eq:Smat})
indeed satisf\/ies these relations.
We can solve the recursive def\/ining relations for $A_P$ and we f\/ind,
with a particular choice of normalisation for~$A_{\rm Id}$, the
following explicit form, for any $P\in\fS_m$
\begin{gather*}
 A_{P}=\prod_{j<k}
\left(1+\frac{1}{2s}~\frac{(e^{ik_{Pj}}-1)(e^{ik_{Pk}}-1)}{e^{ik_{Pj}}-e^{ik_{Pk}}}\right).
\end{gather*}

This scattering matrix becomes after the change of variables
$u=\frac{\lambda+is}{\lambda-is}$
and $v=\frac{\mu+is}{\mu-is}$
\begin{gather*}
 \sigma(\lambda,\mu)=\frac{\lambda-\mu-i}
{\lambda-\mu+i} .
\end{gather*}
Let us remark that after this change of variables the scattering
matrix $\sigma(\lambda,\mu)$ does not depend
on the value of spin and is similar to the one obtained for $s=\half$.

{\bf Projection on
$\boldsymbol{|x_1,\dots,\underbrace{x_i,\ldots,x_i}_{m_1},
\underbrace{x_{i}+1,\ldots,x_{i}+1}_{m_2},\dots,x_m\rangle}$.}
One can compute
\begin{gather}
 \Big[P_{m_{1},m_{2}}(S_{i},\dots,S_{i+m_{1}-1};
S_{i+m_{1}},\dots,S_{i+m_{1}+m_{2}-1})\nonumber\\
\qquad{} +P_{m_{2},m_{1}}\big(
S_{i+m_{1}+m_{2}-1}^{-1},\dots,S_{i+m_{1}}^{-1};
S_{i+m_{1}-1}^{-1},\dots,S_{i}^{-1}\big)\Big]
\nonumber\\
 \qquad{} \times a(\dots,x_i,\dots,x_i,
x_i+1,\dots,x_i+1,\dots)=0,
\label{eq:poly}
\end{gather}
where
\begin{gather*}
 P_{m_1,m_2}(\mathbf{y};\mathbf{z})  =
\sum_{n=1}^{m_2}
\alpha_n\alpha_{m_2-n}\alpha_{m_1}\beta_{0,m_2}^{n}
z_{m_2-n+1}\cdots z_{m_2}\\
\hphantom{P_{m_1,m_2}(\mathbf{y};\mathbf{z})  =}{}
+\sum_{n=1}^{M_1}\alpha_{m_1-n}\alpha_{m_2+n}\beta_{m_1,m_2}^{-n}
y_{m_1-n+1}\cdots y_{m_1}
\nonumber \\
\hphantom{P_{m_1,m_2}(\mathbf{y};\mathbf{z})  =}{}
-\alpha_{m_1}\alpha_{m_2}\sum_{j=1}^{m_1}\!
\left(\frac{\beta_{1,0}^0+\beta_{0,1}^0}{2}+\beta_{1,0}^{-1} y_{j}\right)\!
-\alpha_{m_1}\alpha_{m_2}\sum_{j=1}^{m_2}\!
\left(\frac{\beta_{1,0}^0+\beta_{0,1}^0}{2}+\beta_{1,0}^{-1} z_{j}\right)
\nonumber \\
\hphantom{P_{m_1,m_2}(\mathbf{y};\mathbf{z})  =}{}
+\half\alpha_{m_1}\alpha_{m_2}(\beta_{m_1,m_2}^{0}
+\beta_{0,m_1}^{0}+\beta_{0,m_2}^{0}) ,
\end{gather*}
and we have used the notation (\ref{eq:M1M2}) and
\begin{gather*}
\mathbf{y} = (y_{1},y_{2},\ldots,y_{m_{1}}) \qquad \mbox{with}\qquad
y_{k}=S_{i+k-1} ,\qquad  1\leq k\leq m_{1} ,\\
\mathbf{z} = (z_{1},z_{2},\ldots,z_{m_{2}}) \qquad \mbox{with}\qquad
z_{k}=S_{i+m_{1}+k-1} ,\qquad 1\leq k\leq m_{2} .
\end{gather*}
Relation (\ref{eq:poly}) is implied by (\ref{eq2s}).
The sketch of the proof goes as follows.

First, one can check directly that\footnote{Multiplication on the right
by $a(\dots,x_i,\dots,x_i,
x_i+1,\dots,x_i+1,\dots)$ will be understood during the proof.}
$P_{0,2}(S_{j},S_{j+1})=0$, $\forall\, j$,
since $P_{0,2}(S_j,S_{j+1})$ corresponds to relation~(\ref{eq2s}). The
same is true for
$P_{2,0}(S_{j+1}^{-1},S_j^{-1})$ (after multiplication by
$S_{j+1}^{-1}S_j^{-1}$).

Next, we rewrite (\ref{eq:poly}) as
\begin{gather}
P_{m_1,m_2}(\mathbf{y};\mathbf{z})+
P_{m_2,m_1}(\overline{\mathbf{z}};\overline{\mathbf{y}})=0 ,
\label{eq:poly2}
\end{gather}
where we have def\/ined
\begin{gather*}
\overline{\mathbf{y}} = \big(y_{m_1}^{-1},\ldots,y_{2}^{-1},y_{1}^{-1}\big),
\qquad \mbox{i.e.}\qquad
\overline{y}_{k}=S_{i+m_{1}-k}^{-1} ,\qquad  1\leq k\leq m_{1} ,\\
\overline{\mathbf{z}} = \big(z_{m_2}^{-1},\ldots,z_{2}^{-1},z_{1}^{-1}\big),\qquad  \mbox{i.e.}\qquad
\overline{z}_{k}=S_{i+m_{1}+m_{2}-k}^{-1} ,\qquad  1\leq k\leq m_{2} .
\end{gather*}
These variables are such that if
$P_{0,2}(z_{j},z_{j+1})=0$ $\forall\, j$, then we have also
$P_{0,2}(\overline{z}_{j},\overline{z}_{j+1})=0$. Hence, a property
valid for $P_{m_1,m_2}(\mathbf{y};\mathbf{z})$ will be also valid for
$P_{m_2,m_1}(\overline{\mathbf{z}};\overline{\mathbf{y}})$.

We f\/irst focus on
\begin{gather*}
 P_{0,m}(\mathbf{z})  =
\sum_{n=1}^{m}
\alpha_n \alpha_{m-n} \beta_{0,m}^{n}
z_{m-n+1}\cdots z_{m}
-\alpha_{m} \beta_{1,0}^{-1} \sum_{j=1}^{m} z_{j}+
\alpha_{m}\big(\beta_{0,m}^{0}-m \beta_{1,0}^{0}\big) .
\end{gather*}
If we suppose that we have variables $z_{j}$ such that
$P_{0,2}(z_{j},z_{j+1})=0$, $\forall\,  j$, then from expression~(\ref{eq:beta-facto}), and after some
calculation, one can show that
\begin{gather*}
z_{1} z_{2}\cdots z_{m}  =  1-m+\sum_{j=1}^m \chi^{(m)}_{j}  z_{j},\\
\chi^{(m)}_{j}  =  (-1)^{m+j} \prod_{k=1}^{m-j}\left(\frac{2s}{k}-1\right)
\prod_{\ell=1}^{j-1}\left(\frac{2s}{\ell}+1\right) .
\end{gather*}
Thus, the polynomial $P_{0,m}(\mathbf{z})$
becomes a linear function of the~$z_{j}$'s.
Looking at the coef\/f\/icient of~$z_{j}$ and at the constant term, one
checks that they identically vanish, so that
$P_{0,m}(\mathbf{z})=0$.

Looking at the general polynomial
$P_{m_1,m_2}(\mathbf{y},\mathbf{z})$ and using the relation
\begin{gather*}
 \alpha_{m_{1}-n} \alpha_{m_{2}+n}\beta_{m_{1},m_{2}}^{-n} =
\alpha_{M_{2}} \alpha_{M_{1}-n}\alpha_{n}\beta_{0,M_{1}}^{n}  ,
\end{gather*}
 one can rewrite it as
\begin{gather*}
P_{m_1,m_2}(\mathbf{y},\mathbf{z})  =
\alpha_{m_{1}}\,P_{0,m_2}(\mathbf{z})+
\alpha_{M_{2}}\,P_{0,M_1}(y_{m_1-M_1+1},\ldots,y_{m_1})
+\alpha_{M_{1}}\,\alpha_{M_{2}}\,R_{m_1,m_2}(\mathbf{y}),
 \\
R_{m_1,m_2}(\mathbf{y})  =  -\sum_{j=1}^{m_{1}-M_{1}}
\left(\frac{\beta_{1,0}^0+\beta_{0,1}^0}{2}+\beta_{1,0}^{-1} y_{j}\right)
+\half\big(\beta_{0,M_2}^{0}-\beta_{M_1,0}^{0}
+\beta_{0,m_1}^{0}-\beta_{0,m_2}^{0}\big) .
\end{gather*}
Thus, to prove relation (\ref{eq:poly2}), it is enough to show that
\begin{gather}
R_{m_1,m_2}(\mathbf{y})+R_{m_2,m_1}(\overline{\mathbf{z}})=0 .
\label{eq:polR}
\end{gather}
Using the expression of $M_{1}$ and $M_{2}$, see (\ref{eq:M1M2}),
it is easy to see that
\begin{gather*}
m_{1}-M_{1}=m_{2}-M_{2}\equiv \fm_{12}\geq0 .
\end{gather*}
Two cases have to be distinguished: $\fm_{12}=0$ or $\fm_{12}>0$. In
the f\/irst case, relation (\ref{eq:polR}) is trivially satisf\/ied.
In the second case, equation (\ref{eq:polR}) can be rewritten as
\begin{gather*}
\sum_{j=1}^{\fm_{12}} \big(y_{j} + \overline{z}_{j}\big)
= 2 \fm_{12} ,
\end{gather*}
which is obeyed if
\begin{gather*}
S_{i+j-1}+S_{i+2s+j-1}^{-1}=2 .
\end{gather*}
To prove this last relation, we use recursively (\ref{eq2s}) to show
\begin{gather*}
S_{k}\,S_{k+\ell}+1+\left(\frac{2s}{\ell}-1\right)S_{k}-\left(\frac{2s}{\ell}+1\right)S_{k+\ell}=0
  \qquad \forall\, j.
\end{gather*}
Taking $k=i+j-1$, $\ell=2s$ and using $m_{1}+m_{2}=2s+\fm_{12}$ gives the result.

Hence, relation (\ref{eq:poly2}) is satisf\/ied if the variables are such
that $P_{0,2}(z_{j},z_{j+1})=P_{0,2}(y_{j},y_{j+1})=0$, $\forall\, j$.
This ends the proof.

This step concludes the bulk part of the problem, the other possible
equations being fulf\/illed by linearity. It remains to
take into account the periodic boundary condition. It is done through
the following projection.

{\bf Projection on $\boldsymbol{|1,x_2,\dots, x_m\rangle}$.}
As usual, this leads to a constraint on the parameters $k_{j}$. It is
not surprising since these parameters can be interpreted as
momenta: we are quantizing them since we are
on a line with (periodic) boundary conditions. Namely, this leads to
\begin{gather*}
\sum_{P\in \fS_m}A_P\big(
\exp (i(k_{P2}x_2+\cdots+k_{Pm}x_m))\\
\qquad{}
-\exp (i(k_{P1}x_2+k_{P2}x_3+\cdots+k_{P(m-1)}x_m+k_{Pm}L))
\big)=0 .
\end{gather*}
Now, we f\/irst
perform the change of variable in the summation $P\rightarrow PT_1\cdots T_{m-1}$
in the second term of the previous relation. Then, using recursively relation (\ref{eq:cS})
and projecting on independent exponential functions, we get
the quantization of the momenta via the so-called Bethe equations
\begin{gather}
e^{iLk_j}=\prod_{\ell\neq j}\sigma\big(e^{ik_\ell},e^{ik_j}\big) \qquad \mbox{for}\quad
j=1,2,\dots, m  .
\label{eq:BAE1}
\end{gather}
Since these equations express the periodicity of the chain, they are
equivalent to the ones obtained through projection on
$|x_1,\dots, x_{m-1},L\rangle$ (as it can be checked explicitly).
Thus, we do not have any new independent equations through
projections, and the eigenvalue problem has been solved (up to the
resolution of the Bethe equations).

Note that using the change of variables (\ref{eq:chgV}) and the expression
(\ref{eq:Smat}) for the scattering matrix, equations (\ref{eq:BAE1})
can be rewritten as
\begin{gather*}
\frac{\lambda_j+is}{\lambda_j-is}=-\prod_{\ell=1}^m
\frac{\lambda_\ell-\lambda_j-i}{\lambda_\ell-\lambda_j+i} .
\end{gather*}
One recognizes the usual Bethe equations of the spin $s$
 chain \cite{ABAs,spin-s}.

\subsection*{Action of $\boldsymbol{su(2)}$ generators}
Since the $su(2)$ generators commute with the Hamiltonian, from any
eigenfunction $\Psi_{m}$, one can construct (possibly) new
eigenfunctions by application of $\cS^\alpha$, $\alpha=z,\pm$ on
$\Psi_{m}$. As already
mentioned, it is a straightforward calculation to check that
\begin{gather*}
\cS^z\,\Psi_{m} = (Ls-m) \Psi_{m} .
\end{gather*}
Moreover, it is part of the ansatz to suppose that the eigenvector
$\Psi_{m}$ is a highest weight vector of the $su(2)$ symmetry algebra,
\begin{gather*}
\cS^+ \Psi_{m} = 0 .
\end{gather*}
Let us stress that for $\Psi_{m}$ to be an eigenvector, one has to
assume that the rapidities $\lambda_{j}$ have to obey the Bethe
equations. In the same way, $\Psi_{m}$ is a highest weight vector only
when the Bethe equations are fulf\/illed.
In the context of coordinate Bethe ansatz, there exists no general
proof (for generic spin $s$) of it (at least to our knowledge).
Note however that for spin $\half$, the proof was given in~\cite{Gaudbook}.
Nevertheless, one can check the highest weight property on dif\/ferent cases, and we
illustrate it below by the calculation of $\cS^+ \Psi_{1}$, $\cS^+ \Psi_{2}$
and $\cS^+ \Psi_{3}$. We also show on the last example where the
proof used by Gaudin does not work anymore for spin $s>\half$.

The $\Psi_{m}$ vectors should be also related to the ones obtained through
algebraic Bethe ansatz (ABA). Such a correspondence, for the case of spin $\half$, has
been done in \cite{Ess} using
an iteration trick based on the comultiplication
\cite{IzKo,IzKoRe}.
Let us note that in \cite{Ess} they used the relation $(T_{12})^2=0$ which is not true
anymore for $s>\half$.
Their proof must be generalized to apply in our case. Let us also notice the other
method using the Drinfel'd twist \cite{Ovchi}. Moreover,
since it is known that the ABA construction leads to $su(2)$ highest weight
vectors, and assuming the same property for the coordinate Bethe
approach, it is clear that the two methods should lead to the same
vectors, up to a normalisation.

For instance, considering $\Psi_{1}$, its ABA
``counterpart'' takes the form
\begin{gather*}
\Phi_{1} = \sum_{x=1}^L T^{(1)}_{11}(\lambda_{1})\cdots T^{(x-1)}_{11}(\lambda_{1})
T^{(x)}_{12}(\lambda_{1})T^{(x+1)}_{22}(\lambda_{1})\cdots T^{(L)}_{22}(\lambda_{1})
 |\varnothing\rangle ,
\end{gather*}
where $T^{(j)}(\lambda)$ is the representation of the monodromy matrix at
site $j$:
\begin{gather*}
T^{(j)}(\lambda) = \frac{1}{\lambda-is}\left(\begin{array}{cc}
\lambda+i\, \fs^z_{j} &  i \fs^-_{j}\\
i\fs^+_{j} & \lambda-i\,\fs^z_{j}
\end{array}\right)
\end{gather*}
and $\lambda_{1}$ is the Bethe parameter. This leads to
\begin{gather*}
\Phi_{1} = \frac{i}{\lambda_1+is} \sum_{x=1}^L
\left(\frac{\lambda_{1}+is}{\lambda_{1}-is}\right)^x
\fs^-_{x} |\varnothing\rangle ,
\end{gather*}
that has to be compared with
\begin{gather*}
\Psi_{1} = \sum_{x=1}^{L} e^{ik_{1}x} |x\rangle= \sum_{x=1}^{L} e^{ik_{1}x}
e^-_{x}\,|\varnothing\rangle .
\end{gather*}
Using the change of variable (\ref{eq:chgV}), it is clear that, apart from a
normalisation factor, the two vectors are equal.

{\bf Calculation of $\boldsymbol{\cS^+ \Psi_{1}}$ and $\boldsymbol{\cS^+ \Psi_{2}}$.}
A direct calculation leads to
\begin{gather*}
\cS^+ \Psi_{1} = 2s \frac{y}{1-y}\big(1-y^L\big) |\varnothing\rangle\qquad
\mbox{with} \qquad y=e^{ik_{1}},
\end{gather*}
which is identically zero using the Bethe equation $y^L=1$. Hence,
$\Psi_{1}$ is indeed a highest weight vector for the $su(2)$ symmetry.

In the same way, one can compute
\begin{gather*}
\cS^+ \Psi_{2} = \sum_{P\in\fS_{2}}A_{P}(k_{1},k_{2})\\
\phantom{\cS^+ \Psi_{2} =}{} \times
\left\{ \sum_{x=1}^{L} (2s-1) e^{i(k_{1}+k_{2})x} |x\rangle
+\sum_{1\leq x_{1}<x_{2}\leq L} 2s\,e^{i(k_{P1}x_{1}+k_{P2}x_{2})}
 \big(|x_{1}\rangle+|x_{2}\rangle\big)\right\} .
\end{gather*}
Using the relation
\begin{gather*}
A_{T_{1}}(k_{1},k_{2})=\sigma(k_{1},k_{2}) A_{\rm Id}(k_{1},k_{2})
\end{gather*}
and the normalisation $A_{\rm Id}(k_{1},k_{2})=1$, one gets
\begin{gather*}
\cS^+ \Psi_{2}  =  \sum_{x=1}^{L}\Bigg\{
(2s-1)(y_{1}y_{2})^x \big(1+\sigma(y_{1},y_{2})\big)+
2s\Bigg[\frac{y_{2}^{x+1}-y_{2}^{L+1}}{1-y_{2}}y_{1}^x
\\
\phantom{\cS^+ \Psi_{2}  =}{} +\sigma(y_{1},y_{2}) \frac{y_{1}^{x+1}-y_{1}^{L+1}}{1-y_{1}}y_{2}^x
+\frac{y_{1}-y_{1}^{x}}{1-y_{1}}y_{2}^x
+\sigma(y_{1},y_{2}) \frac{y_{2}-y_{2}^{x}}{1-y_{2}}y_{1}^x
\Bigg]\Bigg\} |x\rangle ,
\end{gather*}
where $y_{j}=e^{i k_{j}}$, $j=1,2$.

Now, from the Bethe equations
\begin{gather*}
y_{1}^{L}=\sigma(y_{1},y_{2}) \qquad \mbox{and} \qquad
y_{2}^{L}=\sigma(y_{2},y_{1}) ,
\end{gather*}
one simplif\/ies it as
\begin{gather*}
\cS^+ \Psi_{2}  =  \sum_{x=1}^{L}(y_{1}y_{2})^x\Bigg\{
2s-1+
2s\left(\frac{y_{2}}{1-y_{2}}-\frac{1}{1-y_{1}}\right)
\\
 \phantom{\cS^+ \Psi_{2}  =}{} +\sigma(y_{1},y_{2}) \Bigg[2s-1+
2s\left(\frac{y_{1}}{1-y_{1}}-\frac{1}{1-y_{2}}\right)
\Bigg]\Bigg\} |x\rangle .
\end{gather*}
Finally, the form of the scattering matrix $\sigma$ ensures that the
quantity within brackets $\{\cdots\}$ vanishes.

{\bf Calculation of $\boldsymbol{\cS^+ \Psi_{3}}$.}
Performing the same kind of calculation on $\Psi_{3}$, we get
\begin{gather*}
\cS^+ \Psi_{3}  =  \sum_{P\in\fS_{3}}A_{P}(\mathbf{k})
\Bigg\{
\sum_{x=1}^{L} (y_{1} y_{2})^{x} \Bigg[ (2s-2) y_{3}^x
+2s \frac{y_{3}^{x+1}-y_{3}^{L+1}}{1-y_{3}}
\\
\phantom{\cS^+ \Psi_{3}  =}{}
+2s
\sigma_{23} \sigma_{13}
\frac{(y_{3}-y_{3}^{x})}{1-y_{3}}\Bigg]
|x,x\rangle
+\sum_{1\leq x_{1}<x_{2}\leq L} y_{1}^{x_{1}} y_{2}^{x_{2}}\Bigg[
(2s-1) \sigma_{23} y_{3}^{x_{1}}+ (2s-1) y_{3}^{x_{2}}
\\
\phantom{\cS^+ \Psi_{3}  =}{}
+2s \frac{y_{3}^{x_{2}+1}-y_{3}^{L+1}}{1-y_{3}} +2s
\sigma_{23} \sigma_{13}
\frac{y_{3}-y_{3}^{x_{1}}}{1-y_{3}} +2s
\sigma_{23}
\frac{y_{3}^{x_{1}+1}-y_{3}^{x_{2}}}{1-y_{3}}\Bigg]
 |x_{1},x_{2}\rangle  \Bigg\} ,
\end{gather*}
where\footnote{Let us stress the dependence on $P\in\fS_{3}$ in the
def\/inition of $y_{j}$: it is used below.}
\begin{gather*}
y_{j}=e^{i k_{Pj}} ,\quad  j=1,2,3
\qquad \mbox{and} \qquad \sigma_{j\ell}=\sigma(y_{j},y_{\ell}) ,\quad  1\leq j\neq \ell\leq3 .
\end{gather*}
After use of the Bethe equation,
$y_{3}^L = \sigma_{23} \sigma_{13}$,
it can be recasted as
\begin{gather*}
\cS^+ \Psi_{3}  =  \sum_{P\in\fS_{3}}A_{P}(\mathbf{k})
\Bigg\{
\sum_{x=1}^{L} (y_{1} y_{2} y_{3})^{x} \left[
2s-2
+\frac{2s}{1-y_{3}} ( y_{3}-
\sigma_{23} \sigma_{13} )
\right]
|x,x\rangle
\\
\phantom{\cS^+ \Psi_{3}  =}{}
+\sum_{1\leq x_{1}<x_{2}\leq L} y_{1}^{x_{1}} y_{2}^{x_{2}}\Bigg[
\sigma_{23} y_{3}^{x_{1}} \left( 2s-1
+2s \frac{y_{3}}{1-y_{3}}-
\frac{2s}{1-y_{3}} \sigma_{13}
\right)
\\
\phantom{\cS^+ \Psi_{3}  =}{}
+y_{3}^{x_{2}} \left( 2s-1+2s \frac{y_{3}}{1-y_{3}}
- \frac{2s}{1-y_{3}} \sigma_{23}
\right)\Bigg]
 |x_{1},x_{2}\rangle  \Bigg\} .
\end{gather*}
Using the sum on $P$ to relabel the variables $y_{j}$, one can
rewrite this equality as
\begin{gather*}
\cS^+ \Psi_{3}  =  \sum_{P\in\fS_{3}}A_{P}(\mathbf{k})
\Bigg\{
\sum_{x=1}^{L} \frac16 (y_{1} y_{2} y_{3})^{x} \\
\phantom{\cS^+ \Psi_{3}  =}{} \times
\Bigg[
(2s-2)\big(1+\sigma_{12}+\sigma_{23}+\sigma_{12}\sigma_{13}
+\sigma_{23}\sigma_{13}+\sigma_{12}\sigma_{13}\sigma_{23}\big)
\\
\phantom{\cS^+ \Psi_{3}  =}{} +
\frac{2s}{1-y_{3}}\big(y_{3}(1-\sigma_{12})-
\sigma_{23}\sigma_{13}-\sigma_{12}\sigma_{13}\sigma_{23}\big)\\
\phantom{\cS^+ \Psi_{3}  =}{}
+
\frac{2s}{1-y_{2}}\big(y_{2}\sigma_{23}(1-\sigma_{13})-
\sigma_{12}-\sigma_{12}\sigma_{13}\big)
\\
\phantom{\cS^+ \Psi_{3}  =}{}
+
\frac{2s}{1-y_{1}}\big(y_{1}\sigma_{12}\sigma_{13}(1-\sigma_{23})-
1-\sigma_{23}\big)
\Bigg]
|x,x\rangle
\\
\phantom{\cS^+ \Psi_{3}  =}{}
+\sum_{1\leq x_{1}<x_{2}\leq L} (y_{1}y_{2})^{x_{1}} y_{3}^{x_{2}}\Bigg[
(1+\sigma_{12}) ( 2s-1 )
+2s\left(\frac{y_{2}}{1-y_{2}}- \frac{1}{1-y_{1}}\right)
\\
\phantom{\cS^+ \Psi_{3}  =}{}
+2s \sigma_{12}\left(\frac{y_{1}}{1-y_{1}}-
\frac{1}{1-y_{2}}\right)\Bigg] |x_{1},x_{2}\rangle
\\
\phantom{\cS^+ \Psi_{3}  =}{}
 +\sum_{1\leq x_{1}<x_{2}\leq L} y_{1}^{x_{1}} (y_{2}y_{3})^{x_{2}}\Bigg[
(1+\sigma_{23}) ( 2s-1 )
+2s\left(\frac{y_{3}}{1-y_{3}}- \frac{1}{1-y_{2}}\right)
\\
\phantom{\cS^+ \Psi_{3}  =}{}
+2s \sigma_{23}\left(\frac{y_{2}}{1-y_{2}}-
\frac{1}{1-y_{3}}\right)\Bigg]
 |x_{1},x_{2}\rangle \Bigg\} .
\end{gather*}
The term inside the square bracket $[\cdots ]$ in factor of
$(y_{1}y_{2})^{x_{1}} y_{3}^{x_{2}}$ on the one hand, and in factor
of $y_{1}^{x_{1}} (y_{2}y_{3})^{x_{2}}$ on the other hand,
identically vanishes. It is in fact the same identity as the one used
to show that $\cS^+ \Psi_{2}=0$. It is also the identity used by
Gaudin~\cite{Gaudbook} to prove, for spin $\half$, that $\Psi_{m}$, $\forall\,
m$, is a highest
weight vector.

When the spin is higher than $\half$, it remains the term in factor
of $|x,x\rangle$, which is a state that does not exist when $s=\half$.
The square bracket in front of $|x,x\rangle$ also identically
vanishes, another identity due to the form of the scattering matrix
$\sigma$, and we
get $\cS^+ \Psi_{3} =0$.

When
$s=1$ this new identity is suf\/f\/icient (together with the one used for
$\cS^+ \Psi_{2}$) to prove that $\Psi_{m}$, $\forall\,
m$, is a highest weight vector. However, for $s>1$, to prove that
$\cS^+ \Psi_{4} =0$, one needs to consider the state
$|x,x,x\rangle$, that we will lead to another identity of the
scattering matrix, and so on: for spin $s$, one needs $2s$ identities
to prove that $\Psi_{m}$, $\forall\, m$, is a highest weight vector.
Hence the dif\/f\/iculty to get a generic proof of it.

\section{Conclusions}\label{conclu}

In previous studies, the eigenfunctions of the spin $s$ chain studied in
this paper
were known thanks to the algebraic Bethe ansatz. This later construction
allows one to compute
the correlation functions~\cite{kit,spinS-fctCorr,spinS-fctCorr2}.
Prior to that computation, the coordinate Bethe ansatz allowed Gaudin~\cite{gau} to
show, for spin~$\half$ chains, orthogonality relations for the Bethe
eigenfunctions, and prove a closure property for these functions.
The explicit form of the
eigenfunctions computed in this note is a f\/irst step toward a
generalisation to spin~$s$ chains. The same method can also be applied
to spin chains associated to higher rank algebras, for which less is
known.

In the same way, the spin 1 XXX chain with open (diagonal) boundaries
has
been studied in~\cite{FiLiUt}: there is no doubt that their results can
be generalized to spin $s$, using the present approach.
The advantage of this method lies in the fact that
we do not need to solve the ref\/lection equation before computing the
spectrum.
We may start with general boundary conditions and f\/ind the ones for
which the method is still
consistent. In this way, the boundaries which keep the model solvable
are classif\/ied.

We also believe that the method presented here can be applied to
solve the XXZ model with higher spin
in the case of periodic boundary conditions. These cases have been
treated through algebraic Bethe ansatz, see e.g.~\cite{XXZ,BR1}. More
interestingly, general XXZ models with
 open boundary conditions can also be treated in this way, see e.g.~\cite{dam} where a f\/irst account has been given.

To conclude, we hope that this paper convinced the reader that the
coordinate Bethe ansatz is a very powerful method and can be applied
to solve a rather large class of integrable
models.

\pdfbookmark[1]{References}{ref}
\LastPageEnding


\begin{thebibliography}{99}

\footnotesize\itemsep=0pt

\bibitem{heis}
Heisenberg W.,
 Zur Theorie des Ferromagnetismus,
{\it Z.~Phys.} {\bf 49} (1928), 619--636.

\bibitem{bethe}
Bethe H.,
Zur Theorie der Metalle. I.~Eigenwerte und Eigenfunktionen der linearen Atomkette,
{\it Z.~Phys.} \textbf{71} (1931), 205--226.

\bibitem{kusk}
Kulish  P.P., Sklyanin E.K.,
Quantum inverse scattering method and the Heisenberg ferromagnet,
\href{http://dx.doi.org/10.1016/0375-9601(79)90365-7}{{\it Phys. Lett.~A}} \textbf{70} (1979), 461--463.

\bibitem{tafa}
Takhtajan  L.A., Faddeev L.D.,
The quantum method of the inverse problem and the Heisenberg XYZ model,
\href{http://dx.doi.org/10.1070/RM1979v034n05ABEH003909}{{\it Russ. Math. Surveys}} \textbf{34} (1979), 11--68.

\bibitem{skl}
Sklyanin E.K.,
Quantum inverse scattering method. Selected topics,
in Quantum Group and Quantum Integrable Systems, Editor Mo-Lin Ge, Singapore,
{\it Nankai Lectures Math. Phys.}, World Sci. Publ., River Edge, NJ, 1992, 63--97,
\href{http://arxiv.org/abs/hep-th/9211111}{hep-th/9211111}.

\bibitem{reshe}
 Vichirko V.I., Reshetikhin N.Yu.,
 Excitation spectrum of the anisotropic generalization of an ${\rm SU}(3)$ magnet,
\href{http://dx.doi.org/10.1007/BF01016823}{{\it Theoret. and Math. Phys.}} \textbf{56} (1983), 805--812. \\
 Reshetikhin N.Yu.,
 A method of functional equations in the theory of exactly solvable quantum systems,
\href{http://www.ams.org/leavingmsn?url=http://dx.doi.org/10.1007/BF00400435}{{\it Lett. Math. Phys.}} \textbf{7} (1983), 205--213.\\
Reshetikhin N.Yu.,
The functional equation method in the theory of exactly soluble quantum systems,
{\it Sov. Phys. JETP} \textbf{57} (1983), 691--696.\\
Reshetikhin N.Yu.,
Integrable models of quantum one-dimensional magnets with $O(n)$ and $Sp(2k)$ sym\-metry,
\href{http://dx.doi.org/10.1007/BF01017501}{{\it Theoret. and Math. Phys.}} \textbf{63} (1985), 555--569.\\
Reshetikhin N.Yu.,
The spectrum of the transfer matrices connected with Kac--Moody algebras,
\href{http://dx.doi.org/10.1007/BF00416853}{{\it Lett. Math. Phys.}} \textbf{14} (1987), 235--246.

\bibitem{krs}
Kulish P.P., Reshetikhin N.Y., Sklyanin E.K.,
Yang--Baxter equation and representation theory.~I,
\href{http://dx.doi.org/10.1007/BF02285311}{{\it Lett. Math. Phys.}} \textbf{5} (1981), 393--403.

\bibitem{fad}
Faddeev L.D.,
How algebraic Bethe ansatz works for integrable model,
 in  Sym\'etries Quantiques  (Les Houches, 1995), Editors A.~Connes, K.~Gawedzki and J.~Zinn-Justin,
{\it Les Houches Summerschool Proceedings}, Vol.~64, North-Holland, Amsterdam, 1998, 149--219,
\href{http://arxiv.org/abs/hep-th/9605187}{hep-th/9605187}.

\bibitem{LimaS} Lima-Santos A.,
Bethe ans\"atze for 19-vertex models,
\href{http://dx.doi.org/10.1088/0305-4470/32/10/004}{{\it J.~Phys.~A: Math. Gen.}} \textbf{32} (1999), 1819--1839,
\mbox{\href{http://arxiv.org/abs/hep-th/9807219}{hep-th/9807219}}.

\bibitem{ABAs}
 Takhtajan L.A.,
 Introduction to algebraic Bethe ansatz,
 in Exactly Solvable Problems in Condensed Matter and Field Theory, Editors  B.S.~Shastry, S.S.~Jha and V.~Singh,
{\it Lecture Notes in Physics}, Vol.~242, Springer, Berlin~-- Heidelberg, 1985, 175--220.


\bibitem{spin-s}
Zamolodchikov A.B., Fateev V.A.,
A model factorized $S$-matrix and an integrable spin-1 Heisenberg ferromagnet,
{\it Soviet J. Nuclear Phys.} \textbf{32} (1980), 298--303. \\ 
 Takhtajan L.A.,
 The picture of low-lying excitations in the isotropic Heisenberg chain of arbitrary spins,
\href{http://dx.doi.org/10.1016/0375-9601(82)90764-2}{{\it Phys. Lett.~A}} \textbf{87} (1982), 479--482.\\
 Babujian H.,
 Exact solution of the isotropic Heisenberg chain with arbitrary spins: thermodynamics of the model,
\href{http://dx.doi.org/10.1016/0550-3213(83)90668-5}{{\it Nuclear Phys.~B}} \textbf{215} (1983), 317--336.

\bibitem{Gaudbook}
Gaudin M.,
La fonction d'onde de Bethe,
 Masson, Paris, 1983.

\bibitem{Ess} Essler F.H.L., Frahm H., G\"ohmann F., Kl\"umper A., Korepin V.E.,
The one-dimensional Hubbard model,
Cambridge University Press, Cambridge, 2005.

\bibitem{IzKo}
Izergin A.G., Korepin V.E.,
The quantum inverse scattering approach to correlation functions,
\href{http://dx.doi.org/10.1007/BF01212350}{{\it Comm. Math. Phys.}} \textbf{94} (1984), 67--97.

\bibitem{IzKoRe}
Izergin A.G., Korepin V.E., Reshetikhin N.Yu.,
Correlation functions in a one-dimensional Bose gas,
\href{http://dx.doi.org/10.1088/0305-4470/20/14/022}{{\it J.~Phys.~A: Math. Gen.}} \textbf{20} (1987), 4799--4822.

\bibitem{Ovchi}
Ovchinnikov A.A.,
Coordinate space wave function from the algebraic Bethe ansatz for the inhomogeneous six-vertex model,
\href{http://dx.doi.org/10.1016/j.physleta.2010.01.022}{{\it Phys. Lett.~A}} \textbf{374} (2010), 1311--1314,
\href{http://arxiv.org/abs/1001.2672}{arXiv:1001.2672}.

\bibitem{kit}
Kitanine N.,
Correlation functions of the higher spin XXX chains,
\href{http://dx.doi.org/10.1088/0305-4470/34/39/314}{{\it J.~Phys.~A: Math. Gen.}} \textbf{34} (2001), 8151--8169,
\href{http://arxiv.org/abs/math-ph/0104016}{math-ph/0104016}.

\bibitem{spinS-fctCorr}
 Castro-Alvaredo O.A., Maillet J.M.,
Form factors of integrable Heisenberg (higher) spin chains,
\href{http://dx.doi.org/10.1088/1751-8113/40/27/004}{{\it J.~Phys.~A: Math. Theor.}} \textbf{40} (2007), 7451--7471,
\href{http://arxiv.org/abs/hep-th/0702186}{hep-th/0702186}.

\bibitem{spinS-fctCorr2}
 Deguchi T., Matsui C.,
 Form factors of integrable higher-spin XXZ chains and the af\/f\/ine quantum-group symmetry,
\href{http://www.ams.org/leavingmsn?url=http://dx.doi.org/10.1016/j.nuclphysb.2009.01.002}{{\it Nuclear Phys.~B}} \textbf{814} (2009), 405--438,
\href{http://arxiv.org/abs/0807.1847}{arXiv:0807.1847}. \\
Deguchi T., Matsui C.,
Correlation functions of the integrable
higher-spin XXX and XXZ chains through the fusion method,
\href{http://dx.doi.org/10.1016/j.nuclphysb.2009.12.030}{{\it Nuclear Phys.~B}} \textbf{831} (2010), 359--407,
\href{http://arxiv.org/abs/0907.0582}{arXiv:0907.0582}.

\bibitem{gau}
Gaudin M.,
Bose gas in one dimension. I.~The closure property of the scattering wavefunctions,
\href{http://dx.doi.org/10.1063/1.1665790}{{\it J. Math. Phys.}} \textbf{12} (1971), 1674--1676.\\
Gaudin M.,
Bose gas in one dimension. II.~Orthogonality of the scattering states,
\href{http://www.ams.org/leavingmsn?url=http://dx.doi.org/10.1063/1.1665791}{{\it J. Math. Phys.}} \textbf{12} (1971), 1677--1680.


\bibitem{FiLiUt}
Fireman E.C., Lima-Santos A., Utiel W.,
Bethe ansatz solution for quantum spin-1 chains with boundary terms,
\href{http://dx.doi.org/10.1016/S0550-3213(02)00027-5}{{\it Nuclear Phys.~B}} {\bf 626} (2002), 435--462,
\href{http://arxiv.org/abs/nlin.SI/0110048}{nlin.SI/0110048}.

\bibitem{XXZ}
 Melo C.S., Martins M.J.,
 Algebraic Bethe ansatz for U(1) invariant integrable models: the method and general results,
\href{http://dx.doi.org/10.1016/j.nuclphysb.2008.07.023}{{\it Nuclear Phys.~B}} \textbf{806} (2009), 567--635,
\href{http://arxiv.org/abs/0806.2404}{arXiv:0806.2404}.\\
 Martins M.J., Melo C.S.,
Algebraic Bethe ansatz for U(1) invariant integrable models: compact and non-compact applications,
\href{http://dx.doi.org/10.1016/j.nuclphysb.2009.04.018}{{\it Nuclear Phys.~B}} \textbf{820} (2009), 620--648,
\href{http://arxiv.org/abs/0902.3476}{arXiv:0902.3476}.

\bibitem{BR1}
Belliard S., Ragoucy E.,
The nested Bethe ansatz for `all' closed spin chains,
\href{http://dx.doi.org/10.1088/1751-8113/41/29/295202}{{\it J.~Phys.~A: Math. Theor.}} \textbf{41} (2008), 295202, 33~pages,
\href{http://arxiv.org/abs/0804.2822}{arXiv:0804.2822}.

\bibitem{dam}
 Cramp\'e N., Ragoucy E., Simon D.,
Eigenvectors of open XXZ and ASEP models for a class of non-diagonal boundary conditions,
\href{http://dx.doi.org/10.1088/1742-5468/2010/11/P11038}{{\it J.~Stat. Mech. Theory Exp.}} \textbf{2010} (2010), no.~11, P11038, 20~pages,
\href{http://arxiv.org/abs/1009.4119}{arXiv:1009.4119}.

\end{thebibliography}
\end{document}